%
\input harvmac.tex
\hfuzz 15pt
\input amssym.def
\input amssym.tex


\input epsf

\font\ninerm=cmr9
\font\eightrm=cmr8



\def\hf{{\litfont {1 \over 2}}}

\def\p{\partial}

\def\a{\alpha}

\def\g{\gamma}

\def\e{\epsilon}

\def\l{\lambda}

\def\G{\Gamma}

\def\L{\lambda_{_{^L}} }

\def\O{\Omega}



  \def\CA {{\cal A}}
  \def\CB {{\cal B}}
  
  \def\CD {{\cal D}}

  \def\CG {{\cal G}}

  \def\CT {{\cal T}}

  \def\CW {{\cal W}}

\def\[{\left[}
\def\]{\right]}
\def\({\left(}
\def\){\right)}
\def\<{\left\langle\,}
\def\>{\,\right\rangle}
 \def\hf{ \frac{1}{2}}


\def\inv{^{-1}}

\def\hf{\textstyle{1\over 2}}

 \def\frac#1#2{ {{\textstyle{#1\over#2}}}}
\def\inv{^{\raise.15ex\hbox{${\scriptscriptstyle -}$}\kern-.05em 1}}

 \def\IN{ \Bbb N}
 \def\IP{\relax{\rm I\kern-.18em P}}

 \def\GL{ C^{\rm Liou}}
  \def\GM{ C^{\rm Matt }}

\def\rb{ \noindent $\bullet$\ \ }

\def\ze{\varepsilon}

\def\zu{\Upsilon}

\def\IR{{ \Bbb R} }
\def\IZ{{ \Bbb Z} }

\def\dC{C\kern-6.5pt I}

\def\CA{{\cal A}}       \def\CB{{\cal B}}       
\def\CD{{\cal D}}              
\def\CG{{\cal G}}

       \def\CT{{\cal T}}       
       \def\CW{{\cal W}}

\def\gb{{\bf b}}
\def\gc{{\bf c}}

\def\muM{\mu_{_M}}

\def\LM{  \l_{_{M}} }


\chardef\tempcat=\the\catcode`\@ \catcode`\@=11
\def\cyracc{\def\u##1{\if \i##1\accent"24 i%
    \else \accent"24 ##1\fi }}
\newfam\cyrfam

%

  \def\np#1#2#3{{Nucl. Phys.} {\CB#1} (#2) #3}
  \def\hepth#1{{hep-th/}#1}

\def\encadremath#1{\vbox{\hrule\hbox{\vrule\kern8pt\vbox{\kern8pt
\hbox{$\displaystyle #1$}\kern8pt} \kern8pt\vrule}\hrule}}

\def\Sx{S_x } \def\IT{{\Bbb T}} \def\zum{{\hat \Upsilon}_b}



  \lref\Witten{ E.~Witten,
Nucl.  Phys.  B 373 (1992) 187, hep-th/9108004.  }

 \lref\SeibergS{ N. Seiberg and D. Shih,
JHEP 0402 (2004) 021, hep-th/0312170.  }

 \lref\KMS{ D. Kutasov, E. Martinec and N. Seiberg,
      Phys.  Lett.  B 276 (1992) 437, \hepth{9111048}.  }

\lref\KP{ I.K. Kostov and V.B. Petkova, Non-Rational 2D Quantum
gravity: Ground Ring versus Loop Gas, in preparation.}

\lref\bershkut{ M. Bershadsky and D. Kutasov,
 Nucl.  Phys.  B 382 (1992) 213, \hepth{9204049}.  }

 \lref\kachru{S. Kachru,
 Mod.  Phys.  Lett.  A7 (1992) 1419, hep-th/9201072.}

\lref\KostovCY{ I.K.~Kostov,
Nucl.  Phys.  B 689 (2004) 3, hep-th/0312301.  }

 \lref\ADEold{ I.K. Kostov, 
 \np{326}{1989}{583}.  }

\lref\KostovCG{ I.K.~Kostov, 
Nucl.  Phys.  B 376 (1992) 539, hep-th/9112059.
}

\lref\DF{Vl.S. Dotsenko and V.A. Fateev,
 Nucl.  Phys.  B 251 [FS13] (1985) 691.}

\lref\Ta{J. Teschner,
Phys.  Lett.  B 363 (1995) 65, hep-th/9507109.  }

  \lref\DiK{ P. Di Francesco and D. Kutasov,
Nucl.  Phys.  B 375 (1992) 119, hep-th/9109005.  }

  \lref\nienhuis{B. Nienhuis, ``Coulomb gas formulation of 2-d phase
  transitions", {\it in} Phase transitions and critical phenomena,
  Vol.  11, ed.  C.C. Domb and J.L. Lebowitz (Academic Press, New
  York, 1987), ch.1.  }

\lref\AndrewsAF{ G.~E.~Andrews, R.~J.~Baxter and P.~J.~Forrester,
J. Stat.  Phys.  35 (1984) 193.
}

\lref\PasquierJC{ V.~Pasquier,
Nucl.  Phys.  B 285 (1987) 162.
} \lref\Pasq{V.~Pasquier, ``Operator content of the ADE lattice
models'', J. Phys.  A {\bf 20}, 5707 (1987).  }

\lref\FodaIN{ O.~Foda and B.~Nienhuis,
Nucl.  Phys.  B 324 (1989) 643.
}

\lref\KostovUH{ I.K.~Kostov, B.~Ponsot and D.~Serban,
Nucl.  Phys.  B 683 (2004) 309, hep-th/0307189.
}
%

\lref\KostovAM{ I.K.~Kostov and M.~Staudacher,
Phys.  Lett.  B 305 (1993) 43, hep-th/9208042.
}

\lref\HiguchiPV{ S.~Higuchi and I.K.~Kostov,
Phys.  Lett.  B 357 (1995) 62, hep-th/9506022.  }

\lref\DavidHJ{ F.~David,
Mod.  Phys.  Lett.  A3 (1988) 1651.
}

\lref\DistlerJT{ J.~Distler and H.~Kawai,
Nucl.  Phys.  B 321 (1989) 509.
}

 \lref\DO{ H. Dorn and H.J. Otto:
Nucl.  Phys.  B 429 (1994) 375, hep-th/9403141.  }

\lref\ZZtp{A.B. Zamolodchikov and Al.B. Zamolodchikov,
Nucl.  Phys.  B 477 (1996) 577, hep-th/9506136.
}

\lref\AlZ{Al.B. Zamolodchikov, On the Three-point Function in
Minimal Liouville Gravity,  \hepth{0505063}.
}

\lref\KazakovPT{ V.A.~Kazakov and I.K.~Kostov,
Nucl.  Phys.  B 386 (1992) 520, hep-th/9205059.  }

 \lref\FZZb{V.~Fateev, A.~B.~Zamolodchikov and A.~B.~Zamolodchikov,
``Boundary Liouville field theory. I: Boundary state and boundary
two-point function'',  hep-th/0001012.
}

\voffset 1cm
\null
\vskip 5mm

\overfullrule=0pt
 
 \rightline{\vbox{\baselineskip12pt\hbox
{SPhT-t05/068}\hbox{}}}

\centerline{
 {\bf Bulk correlation functions in  2D quantum gravity}
\foot{Talk  given
at the International Workshop ``Classical
and Quantum Integrable Systems'',  January 24-28, 2005,  Dubna, Russia.} 
}

 \vskip 7mm
 
  \centerline{
 I.K. Kostov$^{1}$ and V.B. Petkova$^2$
  }

 \vskip 1cm

\centerline{\it $^1$ Service de Physique Th{\'e}orique, CNRS -- URA
2306,} \centerline{\it C.E.A. - Saclay, F-91191 Gif-Sur-Yvette, France
}

   \bigskip

   \centerline{ \it $^2$Institute for Nuclear Research and Nuclear
   Energy, } \centerline{ \it 72 Tsarigradsko Chauss\'ee, 1784 Sofia,
   Bulgaria }

\vskip 1cm

\baselineskip=12pt
   \centerline{\bf Abstract}

\medskip

\noindent {\ninerm We compute bulk 3- and 4-point tachyon correlators
in the 2d Liouville gravity with non-rational matter central charge
$c<1$, following and comparing two approaches.  The continuous CFT
approach exploits the action on the tachyons of the ground ring
generators deformed by  Liouville and matter ``screening charges''.
A by-product general formula for the matter 3-point OPE structure
constants is derived.  We also consider a ``diagonal'' CFT of 2D quantum 
gravity,  in which the degenerate fields are restricted to the diagonal of 
the semi-infinite Kac table. The discrete formulation of the theory is a
generalization of the $ADE$ string theories, in which the target space
is the semi-infinite chain of points.  }

\baselineskip=12pt 

\bigskip\bigskip\bigskip

 \newsec{Introduction and summary}

\noindent The observation that the operator product expansions of the
physical operators in the effective 2d CFT describing the quantum
Liouville gravity reduce, modulo BRST commutators, to simple
``fusion'' relations is an old one \refs{\Witten}.  The ghost number
zero operators were argued to close a ring, the ``ground ring'', which
furthermore preserves the tachyon modules.  It is assumed that in the
rational case it coincides with the fusion ring of the minimal $c<1$
theories, see in particular the recent work in \SeibergS. The action
of the ground ring on tachyon modules was used to derive functional
recursive identities for the tachyon correlation functions
\refs{\KMS,\bershkut,\kachru,\KostovCY}.  

In this work (see \KP\ for a more detailed presentation) we reconsider 
and extend this approach for constructing bulk tachyon correlators.  
We study a non-rational, or quasi-rational,  CFT  of 2D quantum gravity, 
whose effective action is that of a gaussian field perturbed by both 
Liouville and matter ``screening charges''. 

We start with a direct  evaluation of the 3-point function as a
product of Liouville and matter OPE structure constants. For this
purpose we derive an explicit expression, formula (3.3) below, 
 for  the general matter 3-point OPE structure constants. 
 (This result was independently  obtained by Al. Zamolodchikov 
 \AlZ,  with a different normalization of the fields.)  
Then we write difference recurrence equations for the 4-point 
function of tachyons. 

We also consider another  non-rational CFT of 2D quantum gravity,
in which the degenerate fields  along the diagonal of the semi-infinite 
Kac table form a closed algebra. The CFT in question, which we call
`diagonal CFT',  is described by a perturbation with the four
tachyon operators  whose Liouville component has dimension one
and whose matter component has dimension zero. 
 
 A  microscopic realization of this diagonal CFT
 is given by a non-rational extension of the $ADE$ string theories
introduced in \ADEold\ with semi-infinite discrete target space.  The
loop gas representation of the microscopic theory leads to a target
space diagram technique \KostovCG, which allows to calculate
efficiently the tachyon correlation functions
\refs{\KostovAM,\HiguchiPV}.  We find agreement between the ``diagonal''
CFT and the discrete results.

 \newsec{ Non-rational 2D gravity: effective action, local fields,
ground ring}

\bigskip

\noindent The effective action of Euclidean Liouville gravity (taken on
the sphere) is a
perturbation of the gaussian action
   \eqn\actg{\eqalign{ \CA ^{\rm free} & = {1\over 4\pi}\int d^2 x \[
   (\p \phi)^2 + (\p \chi)^2+ (Q \phi + i e_0\chi ) \hat R
   +4({\bf b}\p_{\bar z}{\bf c} +{\bf \bar b} \p_z {\bf\bar
   c})\] }}
%
of the Liouville $\phi$, and matter $\chi$,  fields and a pair of
reparametrization ghosts.  The background charges, $Q={1\over b}+b$
and $e_0={1\over b}-b$
are parametrized by a real $b$, so that the total central charge of
this conformal theory is trivial
  \eqn\totch{ c_{\rm tot}\equiv c_L +c_M+ c_{\rm
  ghosts}=\[13+6(b^2+\frac{1}{b^2})\]+\[13-6(b^2+\frac{1}{b^2})\] -26
  =0.  }
The physical fields, or  the ``on-mass-shell'' tachyons in the
string theory interpretation, are  products of
Liouville and matter vertex operators of total dimension $(1,1)$
\refs{\DavidHJ, \DistlerJT} 
 \eqn\tachyons{ \frac{\g(1-\a^2+e^2)}{\pi} \ e^{2i e\chi}\, e^{2\alpha
 \phi }= \frac{1}{ \pi} \g(\e b^{\e} P)\ e^{i(e_0-P)\chi +(Q-\e
 P)\phi}= \ V_{\a}^{\e}\,, \ \e=\pm 1, }
\eqn\mashell{ e(e-e_0) +\alpha (Q-\alpha ) = 1\ \Rightarrow \ e=\a
-b\,, {\rm or}, \, e=-\a+\frac{1}{b}\,. }
The parameters $P$ and $\e$ in \tachyons\ are interpreted as the
tachyon target space momentum and chirality.  In the ``leg factor''
normalization in \tachyons, $\g(x) =\G(x)/\G(1-x)$. 

The BRST invariant operators  associated  with \tachyons\ are obtained
either by integrating over the world sheet,
or by multiplying with the ghost field ${\bf c \bar{c}}$
of dimension $(-1,-1)$
\eqn\twoforms{\eqalign{\CT_P^{(\pm)} = T_\a^{\pm} \equiv \int
V_{\a}^{\pm} \qquad & {\rm or} \qquad \CW _P^{(\pm)}=W _\a^{\pm}\equiv
{\rm \bf c \bar c}\ V_{\a}^{\pm}.  } }
In the $n$-point tachyon correlators $n-3$ vertex operators
are integrated over the worldsheet and three are placed, as usual, at
arbitrary points, say $0,1 $ and $\infty$. 

The ground ring of BRST invariant operators of zero dimension and
zero ghost number \Witten\ is generated by the two operators
$a_{\pm}(x)=a_{\pm}(z)\,a_{\pm}(\bar z)\,,$
\eqn\bulkGR{ \eqalign{ 
a_-(z)= & \(\gb(z)\gc(z) - b^{-1}
\p_z(\phi(z)+i \chi(z))\) e^{ -b(\phi(z) -i\chi(z))}\cr 
a_+(z)= & \(\gb(z)\gc(z) -b\, 
\p_z(\phi(z)-i \chi(z))\) e^{- b^{-1}(\phi(z)+i\chi(z))}.  
}}
Unlike the tachyons, the operators \bulkGR\ are made of Liouville and
matter vertex operators, both corresponding to degenerate Virasoro
representations.
We will consider deformations of these free field operators determined
by pairs of Liouville and matter ``screening charges'', i.e, by the
interaction actions
  \eqn\deltaS{\eqalign{
   \CA _{\rm int} &= \int \( \mu_{_{L}} e^{ 2b\phi }+ \muM
  e^{-2ib\chi} \) = \L\, \CT^{(+)}_{e_0}+ \LM\, \CT^{(+)}_{Q} \,, \cr
   \tilde \CA_{\rm int} &= \int\( \tilde\mu_{_{L}}
  e^{2\phi/b}+ \tilde \muM e^{2i \chi/b} \) =\tilde \L\,
  \CT^{(-)}_{e_0}+\tilde \LM\, \CT^{(-)}_{-Q}.  
  }
  }
 The  renormalized by the leg factors in \tachyons\ coupling constants 
 are
\eqn\renor{\eqalign{ {\L} \ \ =\ \ \pi \g(b^2)\, \mu ,& \ \ \ \tilde
\L\ \ = \pi \g(\frac{1}{b^{2}})\, \tilde\mu ,\cr \LM =
\pi \g(-b^2)\, \muM,& \ \ \ \tilde \LM = \pi\g(- \frac{1}{b^{2}})\,
\tilde\muM\,.  }}
 The four terms in \deltaS\ describe perturbations that act
separately on the matter and Liouville fields. In such a theory
the 3-point function factorizes into a product of matter and
Liouville components. 
 
One can imagine more general perturbations by integrated tachyon
fields, which affect simultaneously the matter and Liouville
fields. We will study the simplest example of such a
perturbation, which is described by the two Liouville
``screening charges''  as well us by a pair of tachyons related
to the latter by  
a matter charge reflection  $(e,\a) \to (e_0-e,\a)$, namely $(0,
b)\to (e_0, b)$ and $(0, b^{-1})\to (e_0, b^{-1})$. This
perturbation, which we call {\it diagonal perturbation},
is described by the action 
\eqn\deltadiag{\eqalign{ 
\CA_{\rm int}^{\rm diag} & = \L
\big(\CT^{(+)}_{e_0}+ \LM\tilde \LM \, \CT^{(-)}_{-e_0}\big) \ +
\tilde \L \big(\CT_{e_0}^{(-)} + \LM\tilde \LM \,
\CT^{(+)}_{-e_0}\big)\,.  }}
%

\newsec{The tachyon 3-point function as a product of Liouville and
matter OPE constants
}

\noindent The 3-point function in the non-rational CFT of 2D gravity  
described by the perturbation \deltaS\ factorizes to a product of the 
matter and Liouville three-point OPE constants
 \eqn\Gfactor{
 G_3^{\varepsilon_1\varepsilon_2\varepsilon_3}(\a_1,\a_2, \a_3)= \<
 \CW_{P_1}^{(\varepsilon_1)} \CW_{P_2}^{(\varepsilon_2)}
 \CW_{P_3}^{(\varepsilon_3)} \>= { \GL(\alpha _1,\alpha _2,\alpha
 _3)\GM(e_1,e_2,e_3) \over \pi^3\, \prod _{j=1}^3
   \g( \alpha _j^2-e_j^2)}\,.  }
Here $\a_i$ and $e_i$ are solutions of the on-mass-shell condition
\mashell.
\medskip

For the Liouville 3-point constant we take the expression derived  in
\refs{\DO, \ZZtp}:
\eqn\Lc{ \GL(\alpha_1\,,\,\alpha_2\,,\,\alpha_3)=\( \L^{1/b} \,
b^{2e_0}\)^{Q-\alpha _1-\alpha _2-\alpha _3}\, {\zu_b(b)\, \zu_b(2\a
_1)\,\zu_b(2\a _2)\,\zu_b(2\a _3)\,\over \zu_b(\a _{123}-Q)\,\zu_b(\a
_{23}^1)\,\zu_b(\a _{13}^2)\,\zu_b(\a _{12}^3)} }
with notation $\a _{12}^3=\a _1+\a _2-\a _3\,, \a _{123}=\a _1+\a
_2+\a _3$, etc.  This constant is symmetric 
 with respect to
the 
duality transformation $b\to b^{-1}, \ \L\to \L^{1/b^2}=\tilde \L$.

\medskip

For the matter  3-point OPE constant  we obtain the expression
\eqn\mDOZZ{\eqalign{ \GM(e_1\,,\, &e_2\,,\,e_3)=\( \LM^{1/b} \,
b^{2Q}\)^{e_1+e_2+e_3-e_0}\, {\zum(0)\,
\zum(2e_1)\,\zum(2e_2)\,\zum(2e_3)\,\over
\zum(e_{123}-e_0)\,\zum(e_{23}^1)\,\zum(e_{13}^2)\,\zum(e_{12}^3)}\cr
&= { \L^{{1\over b}\(Q-\sum_i\a_i\)} \, \LM^{-{1\over
b}\(e_0-\sum_ie_i\)}\over b^{\sum_i \e_i}\,\prod_{i=1}^{3}
\gamma(b^{\e_i}(Q-2\alpha_i))} \, {1\over
\GL(\alpha_1\,,\,\alpha_2\,,\,\alpha_3)}\,, \ \ \a_i=\e_i\,
e_i+b^{\e_i } }}
invariant under $b \to -1/b\,, \LM \to (\LM )^{-{1/b^2}}=\tilde
\LM$. 
The derivation of \mDOZZ\
repeats the one for the Liouville case in \Ta, where the formula \Lc\
was determined for positive, irrational $b^2$ as the unique (smooth)
solution of a pair of functional relations, see \KP\ for more
details.
The second line of \mDOZZ\ holds for any choice of the three signs $\e_i$ using
the reflection properties of the Liouville OPE constant \Lc.  The function
in the first line is defined as \eqn\mups{ \zum(x):={1\over
\zu_b(x+b)}={1\over \zu_b(-x+{1\over b})}=\zum(e_0-x) ={\hat
\zu}_{1\over b}(-x) }
and satisfies the functional relations \eqn\recupsm{ \zum(x-b) = \g(b
x)\, b^{1-2bx}\, \zum(x)\,, \ \ \zum(x+{1\over b}) = \g(-{1\over b}
x)\, b^{-1-{2\over b}x}\, \zum(x)\,.  }
Its logarithm admits an integral representation as the one for $\log
\zu_b $, with $Q$ replaced by $e_0$ (whence invariant under the change
$b \to -1/b$), which is convergent (for $b>0$) in the strip $ -b < $
Re $ x < {1\over b}$.  The normalization in \mDOZZ\ is fixed by the
choice
\eqn\neut{ \GM(e_1\,,\,e_2\,,\,e_3)=1\,,\ {\rm for} \ \ \sum_i
e_i=e_0\,.  }
For $\sum_i e_i- e_0=mb-{n\over b}\,, $ $ n,m$ non-negative integers,
the expression \mDOZZ\ is finite for generic $b^2$ and $e_i$ and
reproduces, up to the powers $(-\muM)^{m}(-\tilde \muM)^{n}$,
the
3-point Dotsenko-Fateev constant in (B.10) of \DF{}.

Inserting the two expressions \Lc\ and \mDOZZ\ in \Gfactor\ 
we obtain a simple expression for the tachyon 3-point function
\eqn\ttp{\eqalign{ G_3^{\ze_1\ze_2\ze_3 } (\a_1, \a_2, \a_3) & =
{1\over \pi^3\, b^{\ze_1+\ze_2+\ze_3}}\, \L ^{{1\over b}(Q-\sum_i
\a_i)}\, \LM ^{-{1\over b}(e_0-\sum_i e_i)}\,, }}
reproducing an old result, see \DiK\ and references therein.  
The partition function $Z_L(\L,\LM,b)$ is conventionally determined
 identifying its third derivative with respect to $\L$ with
 $-G_3^{+++}(b,b,b)$.

Apart from the power of $\LM$, this expression does not depend on the
presence of matter screening charges.  On the other hand already the
``neutrality'' condition on the matter charges in \neut, being
simultaneously a constraint on the Liouville charges, simplifies
drastically the constant \Lc\ to \ttp.

The ``matter-Liouville product'' formula \ttp{} is valid 
for generic momenta and 
under the normalization assumptions made for the two
constants in the product.  It is however formal, giving $0\times
\infty$, at the singular points of the constants.  By the same reasons,
the simple relation  obtained by combining 
the matter and Liouville functional
relations, cannot be expected to hold in general.  Motivated by these
observations we reconsider the problem of determining the tachyon
3-point function.  It will be determined as the solution of  a
pair of difference equations which will be derived below as part
of the set of functional
identities for the $n$-point tachyon
correlators.  These equations are weaker than the combined matter
plus Liouville functional identities and \ttp{}\ is 
only the simplest of their solutions.  For the normalized 3-point function
$N_{P_1,P_2,P_3}$, defined by 
\eqn\projthree{ \CG ^{\ze_1\ze_2\ze_3 }(P_1\,,\,P_2\,,\,P_3)\, = {\L
^{{1\over 2b} ( \sum_i \ze_iP_i -Q)}\, \LM ^{{1\over 2b} (e_0-
\sum_iP_i )} \over \pi^3\, b^{\ze_1+\ze_2+\ze_3} } \,
N_{P_1,P_2,P_3}\, } 
these equations read \eqn\homrel{ N_{P_1+
b^{\e},P_2,P_3}+N_{P_1- b^{\e},P_2,P_3}= N_{P_1,P_2+ b^{\e}
,P_3}+N_{P_1,P_2- b^{\e} ,P_3}\ , \ \ \ \e=\pm 1\,.  }
We will see later that a possible solution of these equations is
given by the $sl(2)$ ``fusion rules'', i.e., the 
tensor product
 decomposition multiplicities, 
which take values $1$ or $0$. 
In the theory corresponding to the diagonal action \deltadiag\
the factorization to matter$\times$Liouville does not hold and
the 3-point function is determined by an equation of the
same type, but with shifts of the momenta by $e_0$.

 \newsec{The ground ring action on the tachyons }

\noindent A crucial property of the operators $a_\pm$ \bulkGR\ is that
their derivatives $\p_z a_\pm$ and $\p_{\bar z} a^\pm$ are BRST exact:
$\p_z a_- = \{Q_{\rm BRST}, {\rm\bf b}_{-1} a_-\}$.  Therefore, any
amplitude that involves $ a_\pm$ and other BRST invariant operators
does not depend on the position of $ a_\pm$.  This property allows to
write recurrence equations for the tachyon correlation functions using
that the BRST invariant operators $W^\pm_\a$ form a module of the
ground ring up to commutators with the BRST charge \KMS\
\eqn\actaa{\eqalign{ a_{-} W_\a^{-}& =- W_{\a- {b\over 2} }^{-}\,,
\quad a_{+} W_\a^{+} = - W_{\a- {1\over 2b } }^{+} \cr & a_{-}
W_\a^{+} = a_{+} W_\a^{-} =0\,.  }}
Both relations follow from the free field OPE and are deformed in the
presence of integrated tachyon vertex operators.  Thus the second
relation is modified to \refs{\bershkut,\kachru}
\eqn\aminus{ a_- W ^{+}_{\a} T^{+}_{\a_1} = W ^{+}_{\a+\a_1 -{b\over
2}}\,, \ \ a_+ W ^{-}_{\a} T^{-}_{\a_1} = W^{-}_{\a+\a_1-{1\over 2b}}.
}
A particular example of \aminus\ is provided by
$P_1=e_0$, i.e., $\a_1=b\,,$ or $\a_1=1/b$ respectively.  In this case
$T^{\pm}_{\a_1}$ coincide with the  Liouville interaction terms in 
 \deltaS.  Treating
 them as perturbations amounts to modifying the original
ring generators as \KostovCY\
\eqn\newapm{ a_- \to a_- \(1 - \L\, T^+_b+...\)\,,\ \quad a_+ \to
a_+\(1 - \tilde \L\, T^-_{{1/b}}+...\) }
Another deformation of the ring generators \bulkGR , involving the
matter screening charges, corresponds to the choice $\a_1=0$, i.e.,
$P_1=Q$ or $P_1=-Q$ respectively.
Furthermore the action of the ring generators is nontrivial on some
particular double integrals
\eqn\di{\eqalign{ a_{-} \, W_{\a}^{-} \,T_{\a_1}^{+}\, T _{b-\a_1}^{+}
=- W_{\a+{b\over 2}}^{-}\,, \quad a_{+} \, W_{\a}^{+} \,T_{\a_1}^{-}\,
T_{{1\over b}-\a_1}^{-} =- W_{\a+{1\over 2b}}^{+}.  }}
The choices $\a_1=b\, $ in the first and $\a_1= \frac{1}{b}\, $ in the
second relation in \di\ correspond to the combined matter and
Liouville first order perturbations.

Summarizing, the relations \actaa\ get deformed as follows (we keep
the same notation for the fully deformed ring generators):
\eqn\mpd{\eqalign{ a_- \, W^{+}_{\a} &= - \L\, W^{+}_{\a + {b\over 2}
} - \LM W^{+}_{\a - {b\over 2} }\cr a_- \, W_{\a}^{-} & =- W_{\a-
{b\over 2 } }^{-} - \L\LM\, W_{\a + {b\over 2 } }^{-} \, }}
  \eqn\pmd{\eqalign{ a_+ \, W^{-}_{\a}& = - \tilde \L\ W^{-}_{\a+
  {1\over 2 b} } - \tilde\LM W^{-}_{\a- {1\over 2 b} } \cr a_+ \,
  W_{\a}^{+}& = - W_{\a - {1\over 2 b} } ^{+} - \tilde\L\tilde\LM\,
  W_{\a + {1\over 2 b} } ^{+} \,.  }}
The identities \mpd, \pmd\ generalize the OPE relations obtained in
\refs{\KMS,\bershkut,\kachru,\KostovCY}.  The
two terms in each of these relations are in fact the only one
preserving the condition \mashell, out of the four terms in the OPE of
the fundamental matter and Liouville vertex operators in \bulkGR.
There are further generalizations of \di\ with $m$ integrals of
tachyon operators of the same chirality, if the sum of the Liouville
exponents is respectively $\frac{m}{2} b$ and $\frac{m}{2b}$.
When all momenta correspond 
to   screening charges,  matter or Liouville, 
they appear in equal number $k=m/2$. The corresponding OPE
coefficients vanish for $k>1$, implying no new corrections to the two
term action \mpd, \pmd\ of the ring generators. 

This is not the whole story, however, if each of the generators is  
deformed with  all the four terms in \deltaS. For some momenta on the lattice
$kb+l/b,\ k,l\in \IZ$ the OPE relations \mpd, \pmd\   get further 
modified so that terms reversing the given chirality appear. 
 In particular,  restricting to tachyon  momenta labelled by degenerate
matter 
representations, 
\eqn\degen{\eqalign{ &e_0-2e=P=\pm (-mb+n /b)\,, \ m,n\in \IZ_{>0}\,,
\cr }}
the matter 
reflected images of the two terms in the r.h.s. also appear if both
matter screening charges are taken into account along with one of
the Liouville charges\foot{More precisely this holds for \degen\
taken with the plus sign.}.
Alternatively, one can ``deform'' the tachyon basis in the case
of degenerate representations and consider
combinations invariant under matter reflection.

For the operator $a_+a_- $ perturbed by the diagonal interaction
action \deltadiag\ (to be denoted $ A_{+-}$) we obtain a similar
relation with shifts by $e_0/2$, which we shall write in terms of the
momenta $P= \e (Q-2\a)=e_0-2e$: 
\eqn\compra{\eqalign{
 A_{+-}\,\CW_{P}^{(+)}& =\L\, \CW_{P+e_0}^{(+)} +
{\tilde \L}\LM^{-\frac{e_0}{b}}\, \CW_{P-e_0}^{(+)}\cr
A_{+-}\,\CW_{\a}^{(-)}& =\tilde \L\, W_{P+e_0}^{(-)} +
\L\LM^{-\frac{e_0}{b}}\,\CW_{P-e_0}^{(-)}\,.  }}
The diagonal action \deltadiag\ is designed so that to project the four term
action of the product $a_-a_+$,  deformed according to \mpd\ and \pmd, 
 to the two terms in \compra. This relation holds true for generic momenta, 
 as well as when being restricted to  the diagonal
$P= k e_0,  k\in \IZ$, which, with the exception of the point $P=0$,
describes the diagonal degenerate (order operator) fields.

\newsec{Solutions of the  functional  equation for the 3-point function  }

\noindent
Applying \mpd, \pmd\ in a 4-point function with three tachyons we
obtain the functional equations \homrel\ for the tachyon 3-point
functions.  Analogous relation with shifts by $e_0$ follows from
\compra.  We give here some examples solving these equations besides
\ttp{}.

\medskip

Restricting to tachyons labelled by degenerate matter representations
\degen\
%
%
 and setting $\LM=1$ we postulate that the tachyons 
 labelled by the border lines for $m=0$ or $n=0$ vanish.
 Then  one obtains
 as solutions of \homrel\ the product of $sl(2)$ tensor product
 decomposition multiplicities of finite irreps of dimensions $m_i,
 n_i$\foot{In a recent derivation \SeibergS\ of this result in the
 rational $b^2$ case instead of computing OPE coefficients the ground
 ring itself is identified with the fusion ring of the $c<1$ minimal
 models.}
\eqn\sltwopr{ N_{P_1, P_2, P_3}=N_{m_1,m_2,m_3}N_{m_1',m_2',m_3'}\,,
\qquad
P_i=
-m_ib +m'_i\frac{1}{b}\, ,   }
with
\eqn\sltwo{\eqalign{ &N_{m_1,m_2,m_3}= \cases{ 1 & if $\matrix{
&|m_1-m_2|+1\le m_3\le m_1+m_2-1 \cr & {\rm and}\ \ m_1+m_2+m_3= {\rm
odd} }$\ ;\cr 0 & otherwise .} 
}
} 
These multiplicities satisfy the difference identity
\eqn\sltde{
 N_{m_1+ 1,m_2,m_3}-N_{m_1-
1,m_2,m_3}= N_{m_1+m_2,m_3,1}-N_{m_1- m_2, m_3,1}\,.  
}
the r.h.s. of  which indicates the deviation from the naive 
matter-Liouville product functional relation.
 In the diagonal case $m=n$ the equation implied by
\compra\ is solved by one such factor identifying $P_i=\e_i m_i
e_0\,.$

A second  example, now for $P\in\IR$, is given by the expression dual
to \sltwopr, 
\eqn\degbound{\eqalign{ N_{P_1,P_2,P_3} &=
\sum_{m=0}\sum_{n=0} \big(2\sin \pi m e_0 b\, \sin \pi n
\frac{e_0}{b}\big)^2 \, \chi_{P_1}(m,n) \chi_{P_2}(m,n)\,
\chi_{P_3}(m,n) \,, \cr &\chi_P(m,n)={\sin \pi m Pb \over \sin \pi m
e_0 b} {\sin \pi n P/b\over \sin \pi n e_0/b}=\chi_{-P}(m,n)\, }}
in which the degenerate  representations label the  dual
(boundary) variables.  

A solution of the diagonal difference relations 
 is given by the
multiplicity projecting to diagonal matter charges
\eqn\thpm{\eqalign{ &N_{P_1,P_2,P_3}
=\sum_{l=0}^{\infty}\delta(P_1+P_2+P_3 -(2l+1)e_0)\cr &
N_{P_1+e_0,P_2,P_3} -N_{P_1-e_0,P_2,P_3}= \delta(P_1+P_2+P_3)\,.  }}
There are analogous double sum solutions of the non-diagonal
relations \homrel.

 \newsec{Functional equations for the 4-point tachyon amplitudes }

\noindent The integrated tachyon vertex operators in the $n$-point
functions, $n\ge 4$, play the role of screening charges and the OPE
relations as \aminus\ and \di\ imply new channels in the action of the
ring generators besides \mpd, \pmd.  This leads to additional terms
with less than $n$ fields; alternatively these ``contact'' terms are
expected to account for the skipped BRST commutators in the operator
identities \mpd, \pmd.  We write down as an illustration the relation
for the 4-point function $ G^{-+++}_4(\a_1,\a_2,\a_3,\a_4)= \<
W^{-}_{\a_1}\ \ W ^{+}_{\a_2} \ \ \!  T ^{+}_{\a_3}\ \
W^{+}_{\a_4}\>\,$,
  \eqn\rrminus{ \eqalign{ &G_4^{-+++}(\a_1-\frac{b}{2}
  ,\a_2,\a_3,\a_4)+ \L\,\LM\,G_4^{-+++}(\a_1+ \frac{b}{2}
  ,\a_2,\a_3,\a_4) \cr & -\L G_4^{-+++}(\a_1, \a_2+\frac{b}{2}
  ,\a_3,\a_4)- \LM \,G_4^{-+++}(\a_1, \a_2-\frac{b}{2} ,\a_3,\a_4)\cr
  &=- G_3^{-++}(\a_1,\a_2+\a_3-\frac{b}{2} ,\a_4)\cr & +(\L
  \,\delta_{\a_3,0}+\LM \,\delta_{\a_3,b}\, ) G_3^{-++}(\a_1+\frac{b}{
  2},\a_2, \a_4)\,.  }}
The choice of which of the three chirality plus operators is
represented by an integrated vertex should not be essential - this
leads to a set of relations obtained from \rrminus\ by permutations of
$\a_s, s=2,3,4$.  
 The analogous to \rrminus\ identity for the function $
G^{+---}_4$ with reversed chiralities, resulting from \pmd, is
obtained replacing $b \to 1/b\,, \L \to \tilde \L\,, \LM \to \tilde
\LM$.  Similar relations, but with shifts $e_0/2$, are obtained from
the diagonal ring action \compra.

The relations for the 4-point function are expected to hold for
generic tachyon momenta.  Besides the two special contact terms in the
last line in \rrminus, which come from the double integral relation
\di, there are potentially more terms in the quasi-rational case,
which correspond to multiple
integrals generalizing \di.
These integrals  are not of the type in \DF\ and the
 missing information on these contact terms is a main problem.
Furthermore the functional identity \rrminus\ and its dual
correspond each to one  
of the interaction actions in \deltaS\ and they apply to a restricted
class of correlators. Taking into account  the full deformation
as discussed above adds new potential contact terms, including
ones for generic momenta which modify the recurrence identities.

In the absence of matter screening charges, $\sum_i e_i-e_0=0$, the solutions
of the partially deformed ring relation obtained setting $\LM=0$ in
\rrminus, reproduce the 4-point functions found by other means in \DiK.  
Furthermore  the functional relations
admit solutions  with fixed number
of matter screening charges $\sum_i e_i-e_0=mb-n/b$.
%
Another  application of these relations
  is the case of one  degenerate  and
three generic fields.
As an example we take  $P_2=(\frac{1}{b}-(m+1)b)\,,
m\in \IZ_{\ge 0}$ in the matter
degenerate range, imposing the vanishing of the tachyon at the border
value $P_1=\frac{1}{b}$ of \degen.
The equation \rrminus\ is solved
recursively, 
taking as initial condition $G_4^{-+++}(\a_1,b,\a_3,\a_4)=
-\p_{\L}G_3^{-++}(\a_1,\a_3,\a_4)$.
Skipping the overall normalization
the result reads
\eqn\exa{\eqalign{
&\hat G_4^{-+++}(\a_1,a_2=b +\frac{m b}{
2},\a_3,\a_4)=
(m+1)(\sum_{s\ne 1}\alpha_s-Q +\frac{m b}{2})=(m+1)(\sum_{i=1}^4 \a_i-Q -b) 
\cr
&=
(m+1)(\frac{Q}{2}+\frac{mb}{2})-
\hf  \sum_{s\ne 2} \sum_{k_s=0}^m \e_s (P_s -m b +2 k_s b)
}}
and there is an analogous formula for the dual correlator
$G_4^{+---}(a_1,\a_2=\frac{1}{b}+\frac{n }{ 2b}, \a_3,\a_4)$.
\foot{From these fixed chirality formulae one can extract a symmetric
in the  momenta (``local'') correlator, generalizing the non-analytic
expression of \DiK, with physical intermediate momenta in each of the three
channels.}

The situation with the contact terms is simpler 
in  the theory described by the
action \deltadiag.
Consider the case of diagonal $m=n$
degenerate matter
representations with 3-point function given by the
multiplicity \sltwo.
 The solution for the 4-point function has the
structure of a three channel expansion, generalizing a  formula
of \DiK\ in the case of trivial matter.  It reads,
 skipping the universal $\L,\LM$ prefactor
\eqn\fpconj{\eqalign{ & G_4(\a_1,\a_2,\a_3,\a_4)= -
{1\over 2\pi^3 b
e_0}\, \( \sum_{m=0}(N_{m_1,m_2,m} \, (\frac{Q}{3} -m e_0)\,
N_{m,m_3,m_4} + {\rm permutations})\)\cr &= {1\over \pi^3 b e_0}\,
N_{m_1,m_2,m_3,m_4}\(Q+b - \sum_i\a_i - \hf e_0
(N_{m_1,m_2,m_3,m_4}-1)\)
\,, \quad \a_i=\frac{Q}{2}- m_i\frac{e_0}{2}
\,.  }}
Assuming that the largest of the values $m_i$ is say, $m_1$, i.e.,
$m_1\ge m_s\,, s=2,3,4$, the symmetric in the four arguments function
\fpconj{}\ can be identified with a correlator of type $G_4^{-+++}$.  The
contact term is given by a linear combination of the 3-point functions
\eqn\contt{ [N]_{m_1,m_2+m_3,m_4}:= N_{m_1, m_2+m_3, m_4}-
N_{m_1,|m_2-m_3|,m_4} \,, }
which takes the values $0,\pm 1$.
The second term in \contt\ reflects the fact that the fields vanish at
the border $m=0$ of the diagonal  of \degen\
and can be
interpreted as antisymmetric in $P$ combinations so that a tachyon and
its matter (plus Liouville) reflection image are identified.

\newsec{Microscopic  realization of the diagonal  CFT }

\subsec{The SOS model as a discretization of the gaussian matter field}

\noindent
If the matter field is a free gaussian field with a background charge,
{\it i.e.} when $\LM=0$, the 2D gravity can be realized
microscopically as a particular solid-on solid (SOS) model with
complex Boltzmann weights \nienhuis.  The local fluctuation variable
in the SOS model is an integer {\it height} $x\in {\Bbb Z}$ and the
acceptable height configurations are such that the heights of two
nearest-neighbor points are either equal or differ by $\pm 1$.
Therefore each SOS configuration defines a set of domains of equal
height covering the two-dimensional lattice.  The boundaries of the
domains form a pattern of non-intersecting loops on the lattice.  In
this way the SOS model is also described as a {\it loop gas}, {\it
i.e.} as an ensemble of self-avoiding and mutually avoiding loops,
which rise as the boundaries between the domains of equal height.

The loop gas describes a whole class of solvable height models, as the
restricted SOS models (RSOS) \AndrewsAF\ and their $ADE$
generalizations \PasquierJC, in which the target space $\IT $, in
which the height variable takes its values, is the ensemble of the
nodes of a simply-laced ($ADE$-type) Dynkin graph.  The local
Boltzmann weights of the height models depend on the ``mass'' of the
loops $M$ and on the components $S=\{ S_x\} _{x\in\IT}$ of an
eigenvector of the adjacency matrix $A=\{A_{x,x'}\}$ of the graph
$\IT$:
\eqn\adjm{ \sum_{x'} A_{x,x'} S_{x'} = 2 \cos (\pi p_0)\, \Sx .  }
For the unitary $ADE$ models this is the Peron-Frobenius (PF) vector.

The weight of each height configuration factorizes to a product of the
weights of the connected domains and the loops representing the domain
boundaries.  The weight of a domain $\CD$ is
\eqn\weightD{ \O_{\CD}(x)= (\Sx )^{2-n}, \quad n = \#\ {\rm boundaries
\ of} \ \CD .  }
In addition, the loops, or the domain boundaries, are weighted by a
factor $\exp(-M\times~{\rm Length})$, similarly to the droplets in the
Ising model.  The sum over heights can be easily performed using the
relation \adjm\ and the result is that each loop acquires a factor
$2\cos(\pi p_0)$.

In the SOS model, the role of the PF vector $S$ is played by
 \eqn\eigvZ{ \Sx = \frac{1}{\sqrt{2}}\, e^{i\pi p_0 x} \qquad (x\in
 \IZ ) }
where $p_0$ is any real number in the interval $[0, 1]$.  It has been
conjectured \KostovCG\ using some earlier arguments of \FodaIN, that
the critical behavior of the SOS-model on a lattice with curvature
defects is described in the continuum limit by a gaussian field $\chi$
theory with action \actg.  The height variable $x$ and the background
momentum $p_0$ are related to the gaussian field $\chi$ and the
background electric charge $e_0= \frac{1}{b} - b$ as \KostovUH
\foot{Here we are considering only the dilute phase of the loop gas.
The correspondence in the dense phase is $p_0= 1-b^2, \ \chi =\pi b x
$.}
    \eqn\chihash{ \eqalign{ p_0 = \frac{1}{b^2}-1, \quad \chi =\pi
    x/b.  } }

 If the SOS model is considered in the ensemble of planar graphs with
 given topology, its continuum limit will be described by the full
 action \actg.  The Liouville coupling in \deltaS\ is controlled by an
 extra factor $e^{-\L\times {\rm Area}}$ in the Boltzmann weights
 \weightD.  In the dilute phase $\L \sim M^2$.
The detailed description of the SOS model coupled to gravity can
be found in \KostovCG.

\subsec{\it The theory with $\LM\ne 0$ as a semi-restricted SOS model (SRSOS)}

\noindent
One can   argue that the microscopic realization of the theory \actg\
deformed by the term \deltadiag\ is given for generic $b$ by a
``semi-restricted'' 
height model
coupled to gravity,  with target space $\IT =
\IZ_{>0}$.  The Boltzmann weights are defined by \weightD, with
	\eqn\SSxx{\Sx = \sqrt{2}\, \sin (\pi p_0 x) \qquad (x\in
	\IZ_{>0}).  }
  The    background charge  and the normalization of the field
are again given by    \chihash, for all values of $b$. We will
see later  that, with this identification, the   four point
functions in the `diagonal' CFT  \deltadiag\ and the SRSOS loop
model coincide up to a numerical factor.

 The order local operators in the SOS and SRSOS models are constructed
 by inserting the wave functions
  \eqn\orderp{\psi_{p}(x) = e^{ i \pi (p-p_0)x} \ \ \ {\rm for\ SOS},
  \quad \psi_{p}(x) = { \sin(\pi p x) \over \sin(\pi p_0 x)} \ \ \
  {\rm for\ SRSOS}.}
    In the continuum limit the operators \orderp\ are described by
    conformal fields with dimensions $ \Delta_{p}= \frac{p^2-
    p_0^2}{4(1+ p_0)}=\frac{(P^2-e_0^2)}{4}$.

    \subsec{Target space diagram technique }
  
\noindent The loop gas representation allows to build a target space
diagram technique for the string path integral, described in
\refs{\KostovCG, \KazakovPT, \HiguchiPV}.  The $n$-point functions are
given by the sum of all possible Feynman diagrams composed by
vertices, propagators, tadpoles and leg factors.  The rules are that
the vertices can be attached either to tadpoles, or to the
propagators, or to leg factors.  It is forbidden to attach directly
two vertices, or propagator with a tadpole.  We summarize the Feynman
rules in Fig. 1, where we drew the extremities of the lines in such a
way that the rules for gluing them come naturally.  All the elements
of the diagram technique depend on two types of quantum numbers: the
periodic target space momentum $p+ 2\equiv p$ and the nonnegative
integers $k$.
 
 \bigskip

 \noindent \rb Propagator $D_{k, k'}(p)$
 \eqn\vertics {\eqalign{ D_{00}(p)&= - (|p|-\hf)(|p|-\frac{3}{2})\cr
 D_{01}(p)&=- \hf (p^2-\frac{1}{4})(|p|-\frac{3}{2})(|p|-\frac{5}{2})
 =D_{10}(p), \ \cr D_{11}(p)&=- \hf
 (p^2-\frac{1}{4})(|p|-\frac{3}{2})(|p|-\frac{5}{2}) \[1+ \frac{1}{3}
 (|p|-\frac{5}{ 2}) (|p|-\frac{7}{ 2})\] , \ \ \ \ etc.  \cr} }

\rb Vertices $V_{k_1,...,k_n}(p_1,...,p_n)$:
\eqn\verticess{\eqalign{ V_{k_1,...,k_n}(p_1,...,p_n)&=
\frac{(k_1...+k_n)!  }{ k_1!...k_n!}\ N_{p_1,...,p_n}\,, \ \cr {\rm
where}& \ \ \ \ N_{p_1,...,p_n} = \sum _{ x\in \IN} S_x^2\,
\prod_{j=1}^n \psi_{ p_j }(x) } }
 
 \noindent \rb External line factors $\G_k(p)$:
\eqn\defpi{ \G_k(p) = \frac{1}{k!} (\hf+p)_k (\hf-p)_k } \noindent \rb
Tadpole $B_k (p) $ : \eqn\tadpls{ \eqalign{ B_0 (p) = B_1 (p) = 0\,,
\qquad B_k (p) = - \delta (p, p_0)\ \G_{k-1}(g) , \quad k=2,3,\dots }
}
where $(a)_n ={\G(a+n)\over \G(a)}$ 
is the Pochhammer symbol.  The propagator, vertices and tadpole are
defined above for the interval $-1<p<1$.  Their definition is extended
to arbitrary values of $p$ by requiring periodicity $p\to p+2$.  The
leg factors are not required to be periodic.

\vbox{
    \epsfxsize=200pt
   \vskip 10pt
  \centerline{
   \epsfbox{ 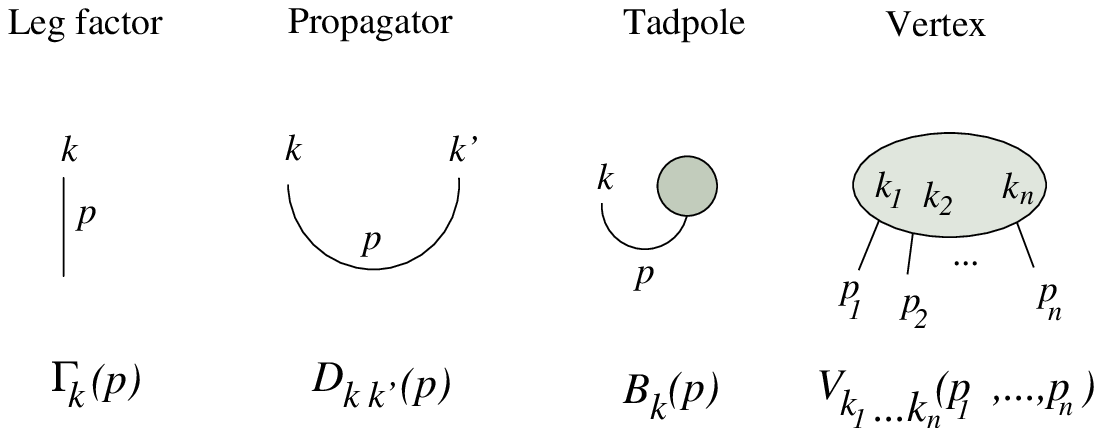   }}
   \vskip 5pt
}{
   \centerline{\eightrm  Fig.1 :  Feynman rules for the
correlation functions    } 
}
    \vskip  10pt

   \bigskip

\subsec{\it General formula for  the 4-point function}

 \noindent The 4-point function is given by the sum of the three
 Feynman diagrams shown in Fig. 2.  Each Feynman diagram stands for the
 sum of terms that differ by permutations of the external legs.

\vbox{
 \epsfxsize=130pt
\vskip 20pt
\centerline{\epsfbox{  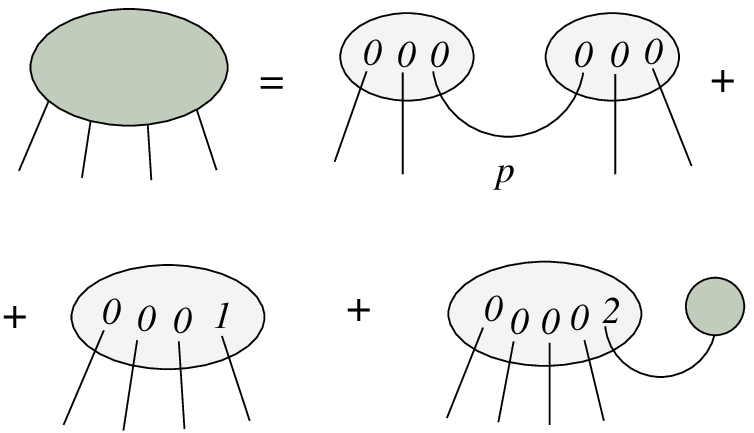  }}
\vskip 5pt
}{
\centerline{\eightrm Fig.2: The diagrams for the 4-point
function      }
}
 \vskip  10pt

\noindent The corresponding analytic expression is
\eqn\ivanb{\eqalign{ G(p_1,p_2,p_3,p_4)&= \int _{-1}^1 dp \[ \G
_0(p_1)\G _0(p_2) N_{p_1p_2 p} D_{00}(p) N_{-pp_3p_4} \G _0(p_3)\G
_0(p_4) + {\rm permutations} \]\cr
&
+\Big[ \G _0(p_1)\G _0(p_2)\G _0(p_3)\G _1(p_4) + {\rm
permutations}\Big] \ N_{p_1p_2p_3p_4} \cr &+ B_2(p_0) \G _0(p_1)\G
_0(p_2)\G _0(p_3)\G _0(p_4) N_{p_0 p_1p_2p_3p_4} }}
Then, using \defpi-\vertics\ we rewrite \ivanb\ as
 \eqn\ibis{\eqalign{ & G(p_1,p_2,p_3,p_4)= \Big[ (1\pm
 p_0)^2-\frac{1}{ 4} + \sum_{s=1}^4\(\frac{1}{ 4}- p_s^2\)
 \Big]N_{p_1,p_2,p_3,p_4} \cr &+ \int _{-1}^1 dp \(
 N_{p_1,p_2,p}N_{-p,p_3,p_4}
 +N_{p_1,p_3,p}N_{-p,p_2,p_4}+N_{p_1,p_4,p}N_{-p ,p_2,p_3}\)
 \(|p|-\hf\)\(|p|-\frac{3}{ 2}\) \cr }}
This formula is valid for both height models, SOS and SRSOS; in
the second case the integration runs over half of the interval. It is
true also for the $ADE$ rational string theories, if the integral over
$p$ is replaced by the corresponding discrete sum.  In the case of
degenerate fields we consider the order operators \orderp\ with $p = m
p_0, \ m= 1,2,3, \dots$ The integral in \ibis\ is replaced by a
summation over the positive integers.  In the 3-point multiplicity in
\vertics\ the summation is replaced by an integral, $x p_0\in [0,2]$,
i.e., the multiplicity is given by the standard integral
representation of \sltwo.  Returning to the normalizations and
notation used in the worldsheet theory, $ P_i= \ze_i bp _i =\ze_i m_i
b p_0\,,\, e_0 = bp_0$, we recover precisely the 4-point formula
\fpconj{}.

 \bigskip
\noindent
{\bf  Acknowledgments}
\smallskip

\noindent {\eightrm 
We thank Al.  Zamolodchikov for several valuable discussions and
for communicating to us  preliminary drafts
 of his forthcoming paper
\AlZ.  We also thank A. Belavin, Vl.  Dotsenko, V. Schomerus and
J.-B. Zuber for the interest in this work and for useful
comments.  This research is supported by the European TMR Network
EUCLID, contract HPRN-CT-2002-00325, and by the Bulgarian
National Council for Scientific Research, grant F-1205/02.
I.K.K. thanks the Institute for Advanced Study, Princeton, and
the Rutgers University for their kind hospitality during part of
this work.  V.B.P. acknowledges the hospitality of Service de
Physique Th\'eorique, CEA-Saclay.  }

\medskip
\listrefs
\bye